\documentclass{jpsj2}
\title{
Spin and Charge Fluctuations and Lattice Effects on Charge Ordering 
in $\alpha$-(BEDT-TTF)$_2$I$_3$ 
}

\author{Satoshi \textsc{Miyashita}$^{1,2}
$\thanks{E-mail address: satoshi@ims.ac.jp} 
and Kenji \textsc{Yonemitsu}$^{2,3}$}

\inst{$^{1}$
Institute for Materials Research, Tohoku University, 
Sendai 980-8577, Japan \\
$^{2}$
Institute for Molecular Science, 
Okazaki 444-8585, Japan \\
$^{3}$
Department of Functional Molecular Science, Graduate University 
for Advanced Studies, Okazaki 444-8585, Japan
}

\abst{
The effects of spin and charge fluctuations and electron-phonon couplings on charge ordering in $\alpha$-(BEDT-TTF)$_2$I$_3$ (BEDT-TTF=bis(ethylenedithio)-tetrathiafulvalence) are investigated theoretically for an anisotropic triangular lattice at 3/4 filling. 
By the exact-diagonalization method, we have calculated the hole density distributions and the modulations of transfer integrals from high-temperature values as a function of electron-phonon coupling strength. 
The results clearly show that the lattice effect on $\alpha$-(BEDT-TTF)$_2$I$_3$ is weak compared with that on $\theta$-(BEDT-TTF)$_2$RbZn(SCN)$_4$, 
as previously found by experiments. 
This finding, which is also consistent with recent mean-field results, is systematically explained by strong-coupling perturbation theory; the effects of spin fluctuations are partially canceled by charge fluctuations in $\alpha$-(BEDT-TTF)$_2$I$_3$. 
}

\kword{organic conductor, charge ordering, extended Hubbard model, 
electron-phonon coupling, exact diagonalization}

\begin{document}
\maketitle

\section{Introduction} 
Two-dimensional (2D) strongly correlated electron systems (SCESs) have attracted the attention of many physical scientists. 
Bis(ethylenedithio)-tetrathiafulvalence (BEDT-TTF) salts, which are typical 2D SCESs, exhibit various interesting phenomena\cite{Ishiguro,Williams} such as metal-insulator (M-I) transitions accompanied with charge ordering, the realization of a spin liquid state,\cite{Shimizu} and superconductivity.\cite{Kanoda} 
In particular, transitions to states with a charge order (CO) have been extensively studied both theoretically\cite{Kino1,Kino2,Seo1,Clay,Seo2,Seo3,Merino} and experimentally.\cite{Bender,HMori1,HMori2,Wojciechowski,NTajima,Takano,Rothaemel,Moroto,Watanabe,Miyagawa,Chiba,HTajima,Yamamoto,Kakiuchi} 
Many theoretical scientists use extended Hubbard models to determine the role played by long-range Coulomb interactions in realizing the CO. 
However, electron-phonon couplings can also have a significant effect because the crystal structure is often altered at the M-I transition. 

$\alpha$-(BEDT-TTF)$_2$I$_3$ (called $\alpha$-I$_3$ for simplicity hereafter) 
is a representative compound that undergoes a CO transition accompanied by structural deformation with decreasing temperature,\cite{Bender,HMori1} which is analogous to $\theta$-(BEDT-TTF)$_2$RbZn(SCN)$_4$ ($\theta$-RbZn hereafter).\cite{Watanabe} 
The ground state of $\alpha$-I$_3$ is an insulator with a horizontal-stripe CO on the $b2'$ and $b3$ bonds (HCO-$t_{b2'}$-$t_{b3}$) as shown in Fig. \ref{f1}(b).\cite{Kakiuchi} 
(Nearly half-filled) Hubbard chains containing hole-rich sites $A$ and $B$ are 
dimerized by the alternation of $b2'$ and $b3$ bonds. 
Accordingly, at the M-I transition, the magnetic susceptibility undergoes a transition into a nonmagnetic state in $\alpha$-I$_3$.\cite{Rothaemel,Moroto} 
The importance of long-range electron-electron interactions is well recognized.
\cite{Seo1} 
The mechanism for stabilizing the HCO has been mainly discussed on the basis of the low-temperature structure.\cite{Kaneko} 
However, electron-phonon couplings give rise to a first-order transition, 
whose discontinuity is sensitive to the crystal structure. 

At the M-I transition, the $\theta$-RbZn and $\alpha$-I$_3$ salts undergo 
large and small molecular rearrangements, respectively. 
Iwai and coworkers have observed the photoinduced melting of the CO in these salts by femtosecond reflection spectroscopy.\cite{Iwai1,Iwai2}
Their photoinduced dynamics are qualitatively different: $\theta$-RbZn exhibits local melting of the CO and ultrafast recovery, while $\alpha$-I$_3$ exhibits critical slowing down. 
Thus, it is important to study why the effects of electron-phonon couplings 
on $\theta$-RbZn are different from those on $\alpha$-I$_3$. 

In our previous paper\cite{Miyashita}, we investigated the effects of electron-phonon couplings on the charge ordering in the $\theta$-RbZn salt. 
Strong-coupling perturbation theory shows that the HCO state is stabilized by transfer modulations that increase linearly with respect to an electron-phonon coupling. 
More specifically, spin fluctuations induce lattice distortions that stabilize the HCO. 
Details are summarized in the Appendix. 

This paper is organized as follows. 
After a brief explanation of the model in the next section, 
we present exact-diagonalization results for hole density distributions and the modulations of transfer integrals in $\S$3. 
We will numerically simulate the lattice effects on the HCO in $\alpha$-I$_3$. 
In $\S$4, we discuss the relation between spin and charge fluctuations and the lattice effects on the HCO using strong-coupling perturbation theory. 
A summary is given in $\S$5. 

\section{Model}
We use the following two-dimensional (2D) 3/4-filled extended Hubbard model with electron-phonon couplings that modulate transfer integrals on the anisotropic triangular lattice shown in Fig. \ref{f1},\cite{Miyashita,Tanaka1,Tanaka2} 
\begin{eqnarray}
 {\cal H} &=& \sum_{<i,j>} 
              \left[
               t_{i,j} \pm \alpha_{i,j} ( u_i - u_j )
              \right] 
              c^\dag_{i\sigma} c_{j\sigma}
           +  U \sum_{i} n_{i\uparrow}n_{i\downarrow}
\nonumber \\
   &&      +  \sum_{<i,j>} V_{i,j} n_i n_j
           +  \sum_{i,j} \frac{K_{i,j}}{2} (u_i - u_j)^2, 
\label{f1}
\end{eqnarray}
where $c_{i\sigma}^\dag$ creates an electron with spin $\sigma$ at site $i$, $n_{i\sigma}=c_{i\sigma}^\dag c_{i\sigma}$, and $n_i=n_{i\uparrow}+n_{i\downarrow}$. 
$t_{i,j}$ is the transfer energy from the $j$th site to the nearest-neighbor $i$th site. 
$U$ represents the on-site Coulomb interaction, and $V_{i,j}=V_c$ or $V_p$ is the intersite Coulomb interaction between the $i$th site and the $j$th site on the vertical bonds or the diagonal bonds, respectively. 
$u_i$ is the $i$th molecular displacement from the high-temperature structure. 
$\alpha_{i,j}$ and $K_{i,j}$ are the corresponding coupling constant and elastic coefficient, respectively. 
Here, we ignore modulations of intersite Coulomb interactions through lattice distortions because the transfer energies are generally much more sensitive to lattice distortions than intersite Coulomb interactions. 
In fact, the experimental findings on BEDT-TTF salts can be basically understood in terms of Peierls-type electron-phonon couplings.\cite{Miyashita,Tanaka1,Tanaka2} 
This is not always the case: in the TTF-CA (TTF-CA=tetrathiafulvalence-chloranil) complex, which undergoes a neutral-ionic transition, for example, modulations of intersite Coulomb interactions play an important role in stabilizing ionic domains.\cite{Kawamoto}

For simplicity, we perform the variable transformations 
\begin{eqnarray}
\begin{array}{lr}
\displaystyle
       \alpha_{i,j} (u_i - u_j) = y_{i,j}, & 
\displaystyle
       \frac{\alpha_{i,j}^2}{K_{i,j}} = s_{i,j}
\;. 
\end{array}
\label{transformation}
\end{eqnarray}
We take the electron volt (eV) as the unit of energy for all variables in the following except for Fig. \ref{f1}. 
The crystal structures of $\alpha$-I$_3$ in the high- and low-temperature phases are shown in Figs. \ref{f1}(a) and \ref{f1}(b), respectively. 
From synchrotron-radiation X-ray crystal-structure analysis,\cite{Kakiuchi} inversion symmetry exists only in the high-temperature phase [Fig. \ref{f1}(a)], where the unit cell consists of two equivalent sites ($A$), a hole-rich site ($B$), and a hole-poor site ($C$). 
In the low-temperature phase below $T_{\rm CO}=135$ K [Fig. \ref{f1}(b)]\cite{Bender}, four nonequivalent sites ($A$, $A'$, $B$, and $C$) exist in the unit cell owing to the symmetry breaking. 

\begin{figure*}[thb]
 \begin{center}
  \begin{tabular}{cc}
   \multicolumn{1}{l}{(a)} & 
   \multicolumn{1}{l}{(b)} \vspace{3mm} \\
   \resizebox{50mm}{!}{\includegraphics{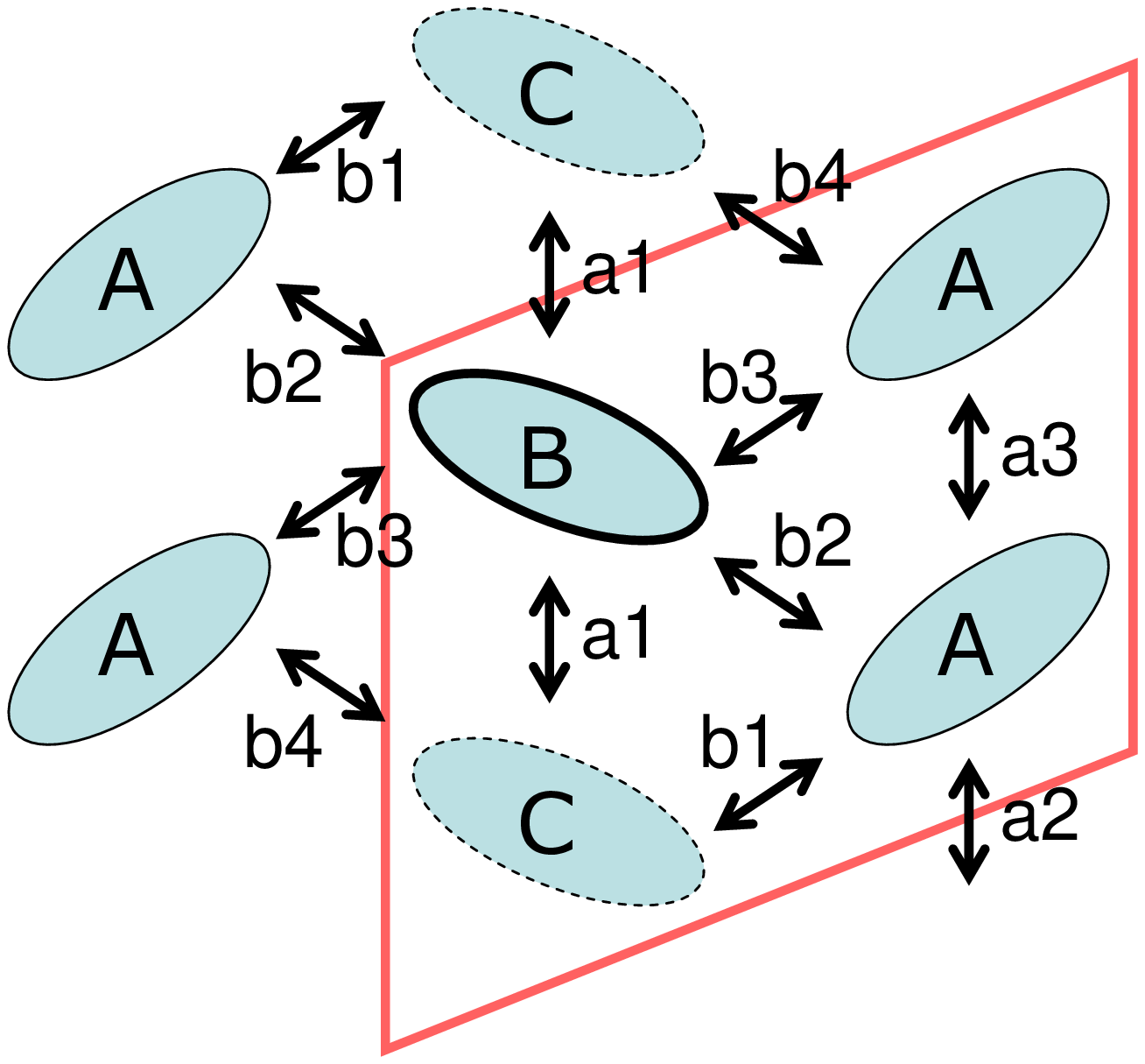}} &
   \resizebox{50mm}{!}{\includegraphics{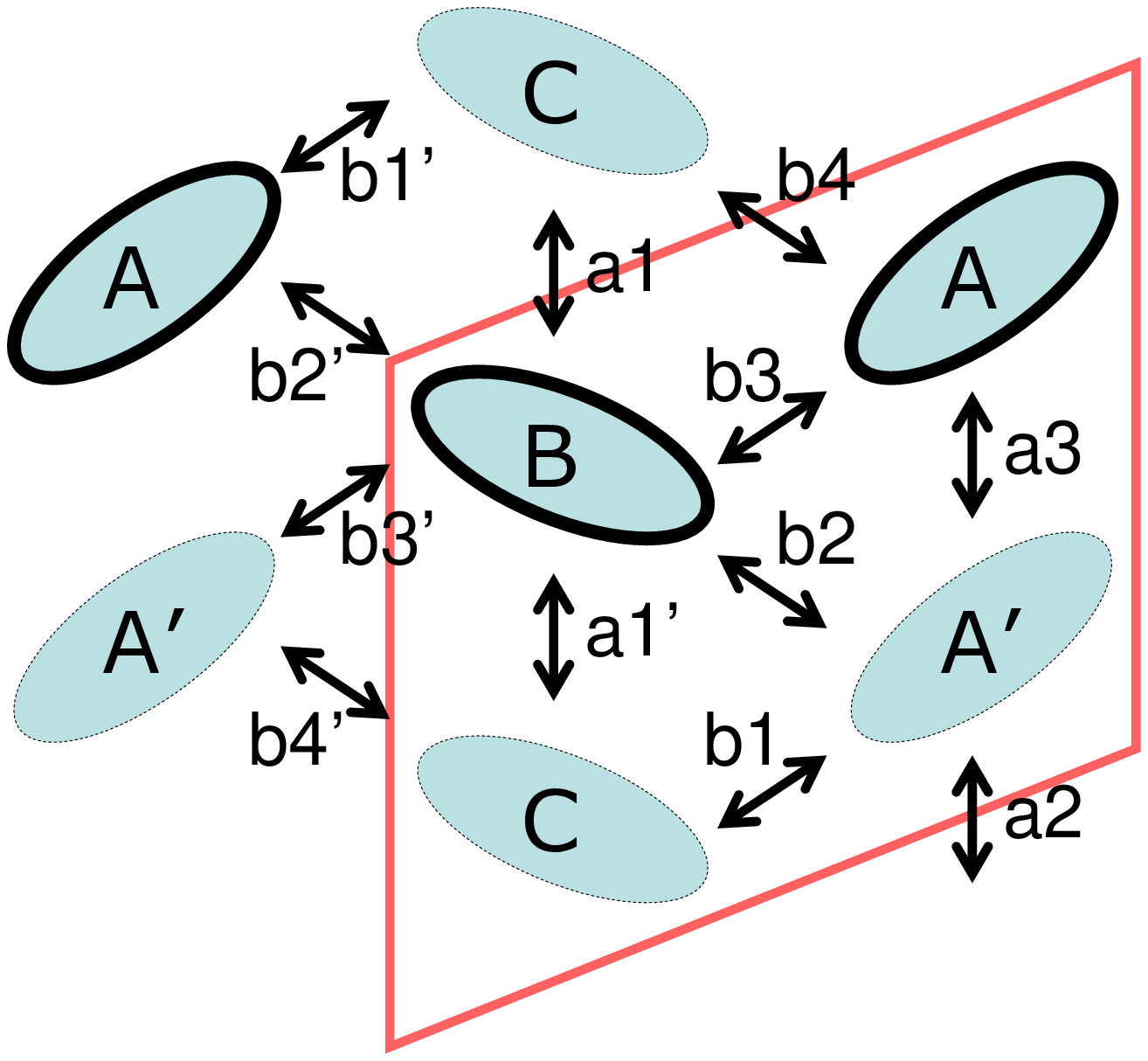}} \\
   $\begin{array}{l}
    t_{a1}^{\rm HT} =-3.50  \\
    t_{a2}^{\rm HT} =-4.61  \\
    t_{a3}^{\rm HT} = 1.81  \\
    t_{b1}^{\rm HT} =12.71  \\
    t_{b2}^{\rm HT} =14.47  \\
    t_{b3}^{\rm HT} = 6.29  \\
    t_{b4}^{\rm HT} = 2.45  \\
   \end{array}$
  & 
   $\begin{array}{ll}
    t_{a1}  =-3.08, &
    t_{a1'} =-4.95  \\
    t_{a2}  =-5.44  & 
                    \\
    t_{a3}  = 3.29  & 
                    \\
    t_{b1}  =12.12, &
    t_{b1'} =16.52  \\
    t_{b2}  =15.77, &
    t_{b2'} =17.73  \\
    t_{b3}  = 6.73, &
    t_{b3'} = 6.56  \\
    t_{b4}  = 0.39, &
    t_{b4'} = 3.23  \\
   \end{array}$
  \\
  \end{tabular}
 \end{center}
\caption{(Color online) 
Anisotropic triangular lattices for 
(a) the high-temperature structure with charge disproportionation and 
(b) the low-temperature structure with HCO-$t_{b2'}$-$t_{b3}$ 
of $\alpha$-I$_3$. 
Transfer integrals estimated by the extended H\"uckel calculation
\cite{Kakiuchi} 
are shown below each figure ($\times$10$^{-2}$ eV).
}
\label{model}
\end{figure*}
For electron-phonon couplings, the ways by which the transfer integrals are modulated from the high-temperature undistorted structure to the low-temperature distorted structure are listed in Table \ref{tbl1}. 
In our calculations below, we ignore the difference between $t_{a2}$ and $t_{a2}^{\rm HT}$ and that between $t_{a3}$ and $t_{a3}^{\rm HT}$, which have been experimentally observed in $\alpha$-I$_3$, because they have an insignificant effects on the CO. 
\begin{table*}[htpb]
\begin{center}
 \begin{tabular}{ll}
   $t_{a1}  = t_{a1}^{\rm HT} + y_{a1}$, &
   $t_{a1'} = t_{a1}^{\rm HT} - y_{a1}$, \\
   $t_{a2}  = t_{a2}^{\rm HT}$,          &
                                         \\
   $t_{a3}  = t_{a3}^{\rm HT}$,          &
                                         \\
   $t_{b1}  = t_{b1}^{\rm HT} - y_{b1}$, &
   $t_{b1'} = t_{b1}^{\rm HT} + y_{b1}$, \\
   $t_{b2}  = t_{b2}^{\rm HT} - y_{b2}$, &
   $t_{b2'} = t_{b2}^{\rm HT} + y_{b2}$, \\
   $t_{b3}  = t_{b3}^{\rm HT} + y_{b3}$, &
   $t_{b3'} = t_{b3}^{\rm HT} - y_{b3}$, \\
   $t_{b4}  = t_{b4}^{\rm HT} - y_{b4}$, &
   $t_{b4'} = t_{b4}^{\rm HT} + y_{b4}$. \\
 \end{tabular}
\end{center}
\caption{
Modulations of transfer integrals from the high-temperature structure 
to the low-temperature structure of $\alpha$-I$_3$.
}
\label{tbl1}
\end{table*}
The signs here are chosen so that $y_{\mu}>0$ corresponds to the experimentally observed modulation. 

In this paper, we apply the values $U=0.7$, $V_c=0.35$, and $V_p=0.30$ to the interaction strengths,\cite{Ducasse,Imamura,TMori1,TMori2} which reasonably reproduce the observed optical conductivity.\cite{Miyashita} 
Typical values for transfer integrals in BEDT-TTF salts are estimated from the extended H\"uckel calculation.\cite{TMori3,TMori4,TMori5} 
From the experimental data\cite{Kakiuchi} shown in Fig. \ref{f1}, we adopt the values of 
$t_{a1}^{\rm HT}=-0.0350$, 
$t_{a2}^{\rm HT}=-0.0461$, 
$t_{a3}^{\rm HT}= 0.0181$, 
$t_{b1}^{\rm HT}= 0.1271$, 
$t_{b2}^{\rm HT}= 0.1447$, 
$t_{b3}^{\rm HT}= 0.0629$, and 
$t_{b4}^{\rm HT}= 0.0245$.

We use the exact-diagonalization method for electrons, regard phonons as classical variables, and apply the Hellmann-Feynman theorem to impose the self-consistency condition on the phonons ($y_{\mu}$), 
\begin{eqnarray}
  \langle
   \frac{{\partial \cal H}}{\partial y_{\mu}}
  \rangle
 = 0 
\;. 
\label{HelFey-theorem}
\end{eqnarray}
\section{Numerical Results}
%
\begin{figure}[thb]
 \begin{center}
   \resizebox{80mm}{!}{\includegraphics{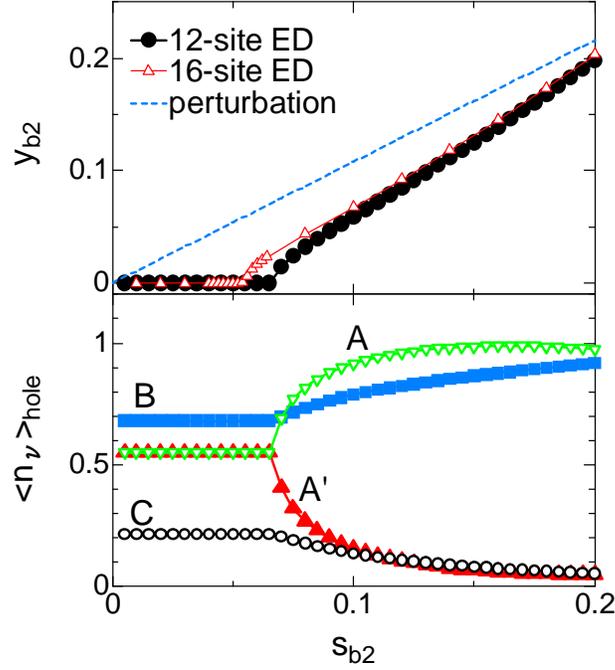}}
 \end{center}
\caption{(Color online) 
Modulations of transfer integrals (upper panel) and hole densities (lower panel) as a function of $s_{b2}$ for $s_{a1}=s_{b1}=s_{b3}=s_{b4}=0$ with $U$=0.7, $V_c$=0.35, and $V_p$=0.30. 
The solid circles (open triangles) in the upper panel represent the 3$\times$4(4$\times$4)-site exact-diagonalization result. 
The dashed line indicates the linear increase obtained from the strong-coupling analysis in $\S$\ref{strong}. 
}
\label{f2}
\end{figure}

Figure \ref{f2} shows the modulations of the transfer integrals on the $b2$ bonds and the hole densities as a function of $s_{b2}$. 
Although the undistorted state with charge disproportionation is stable 
for very small $s_{b2}$, $y_{b2}$ becomes finite above $s_{b2}^{\rm cr} \sim 0.07(0.055)$ for the 4$\times$3$=$12(4$\times$4$=$16)-site cluster and increases almost linearly. 
Then, the HCO-$t_{b2'}$-$t_{b3}$ state (with hole-rich $A$ and $B$ and hole-poor $A'$ and $C$), which is consistent with the experimentally observed CO pattern,\cite{Kakiuchi} is stabilized. 
In the upper panel, the linear increase of $y_{b2}$ derived by perturbation theory in the next section is indicated by the dashed line. 

Because the displacements and bond densities need to be obtained self-consistently, the system is limited to a small size. 
As shown in the upper panel of Fig. \ref{f2}, the difference between the numerical results of the 12-site cluster and those of the 16-site cluster is small. 
Furthermore, these results are explained by strong-coupling perturbation theory, as shown later. 
For these reasons, the system used hereafter is the 12-site cluster with a periodic boundary condition.

%
\begin{figure}[thb]
 \begin{center}
   \resizebox{80mm}{!}{\includegraphics{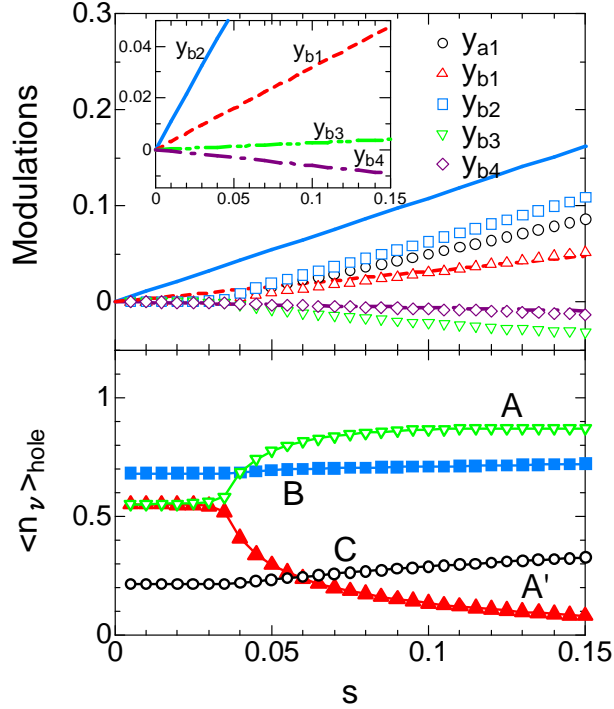}}
 \end{center}
\caption{(Color online) 
Modulations of transfer integrals (upper panel) and hole densities (lower panel) as a function of electron-phonon coupling constant for $s_{a1}=s_{b1}=s_{b2}=s_{b3}=s_{b4}=s$ with $U$=0.7, $V_c$=0.35, and $V_p$=0.30. 
The inset of the upper panel shows the linear increase of the modulations of diagonal bonds, which are obtained by the strong-coupling analysis in $\S$\ref{strong}. 
The linear increases of $y_{b1}$, $y_{b2}$, and $y_{b4}$ are also shown by the dashed line, the solid line, and the dashed-dotted line, respectively, in the upper panel for comparison. 
}
\label{f3}
\end{figure}
Next, we study the effects of the other electron-phonon couplings, which modulate the transfer integrals as shown in Table \ref{tbl1} (assuming $s_{a1}=s_{b1}=s_{b2}=s_{b3}=s_{b4}=s$), on the stability of the HCO-$t_{b2'}$-$t_{b3}$ state. 
The $s$-dependences of the transfer modulations and hole densities are shown in Fig. \ref{f3}. 
The undistorted state with charge disproportionation is unstable 
above $s \sim 0.03$. 
With increasing $s$, all modulations $|y_{\mu}|$ increase almost linearly and 
the CO pattern is always the HCO-$t_{b2'}$-$t_{b3}$ state. 
Contrary to the experimental results, $y_{b3}$ and $y_{b4}$ are negative, 
but their magnitudes are so small that their effects are insignificant. 
More importantly, the HCO-$t_{b2'}$-$t_{b3}$ state is found to be the most stable even if all the bonds are allowed to be distorted. 
We show the linear increase of modulations except for the $a1$ bonds, which are expected to be undistorted from the strong-coupling analysis in $\S$\ref{strong}, in the inset of the upper panel. 
Here also, $y_{b2}$ has the largest value and is comparable to the numerical result.  

\begin{figure}[thb]
 \begin{center}
   \resizebox{80mm}{!}{\includegraphics{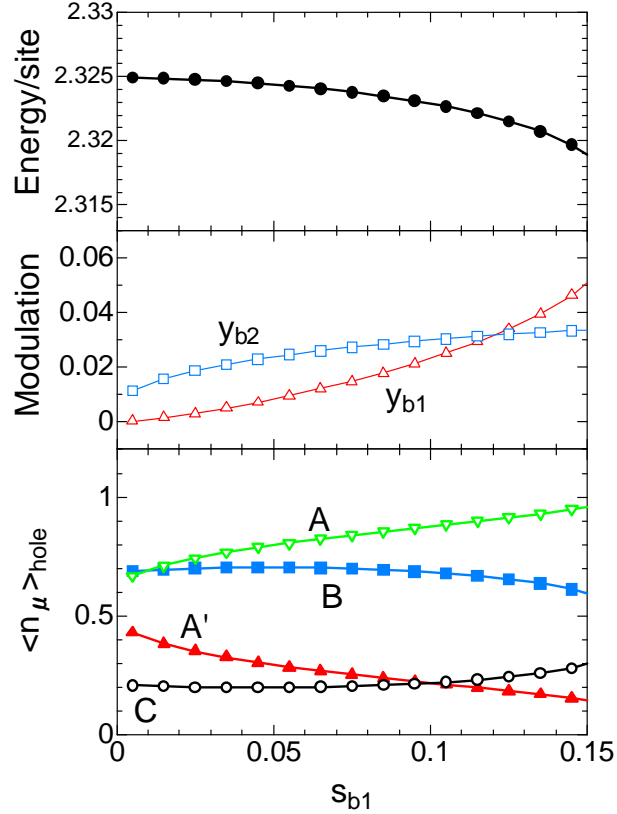}}
 \end{center}
\caption{(Color online) 
$s_{b1}$-dependence of total energy per site (upper panel), modulations of transfer integrals (middle panel), and hole densities (lower panel) with $U=0.7$, $V_c=0.35$, $V_p$=0.30, and $s_{b2}$ fixed at 0.068.
}
\label{f4}
\end{figure}
Let us now consider the effect of $s_{b1}$ under the condition that the HCO-$t_{b2'}$-$t_{b3}$ state is stable. 
Figure \ref{f4} shows the $s_{b1}$-dependence of the ground-state energy per site, the modulations ($y_{b1}$, $y_{b2}$), and the hole densities with $s_{b2}=0.068$. 
The HCO-$t_{b2'}$-$t_{b3}$ state remains stable up to at least $s_{b1}=0.15$ because the ground-state energy is lowered by $s_{b1}$. 
Of course, when $s_{b1}$ is extremely large, the other horizontal-stripe pattern (HCO-$t_{b1'}$-$t_{b4}$) is favored.\cite{Tanaka2} 

%
\begin{figure}[thb]
 \begin{center}
   \resizebox{80mm}{!}{\includegraphics{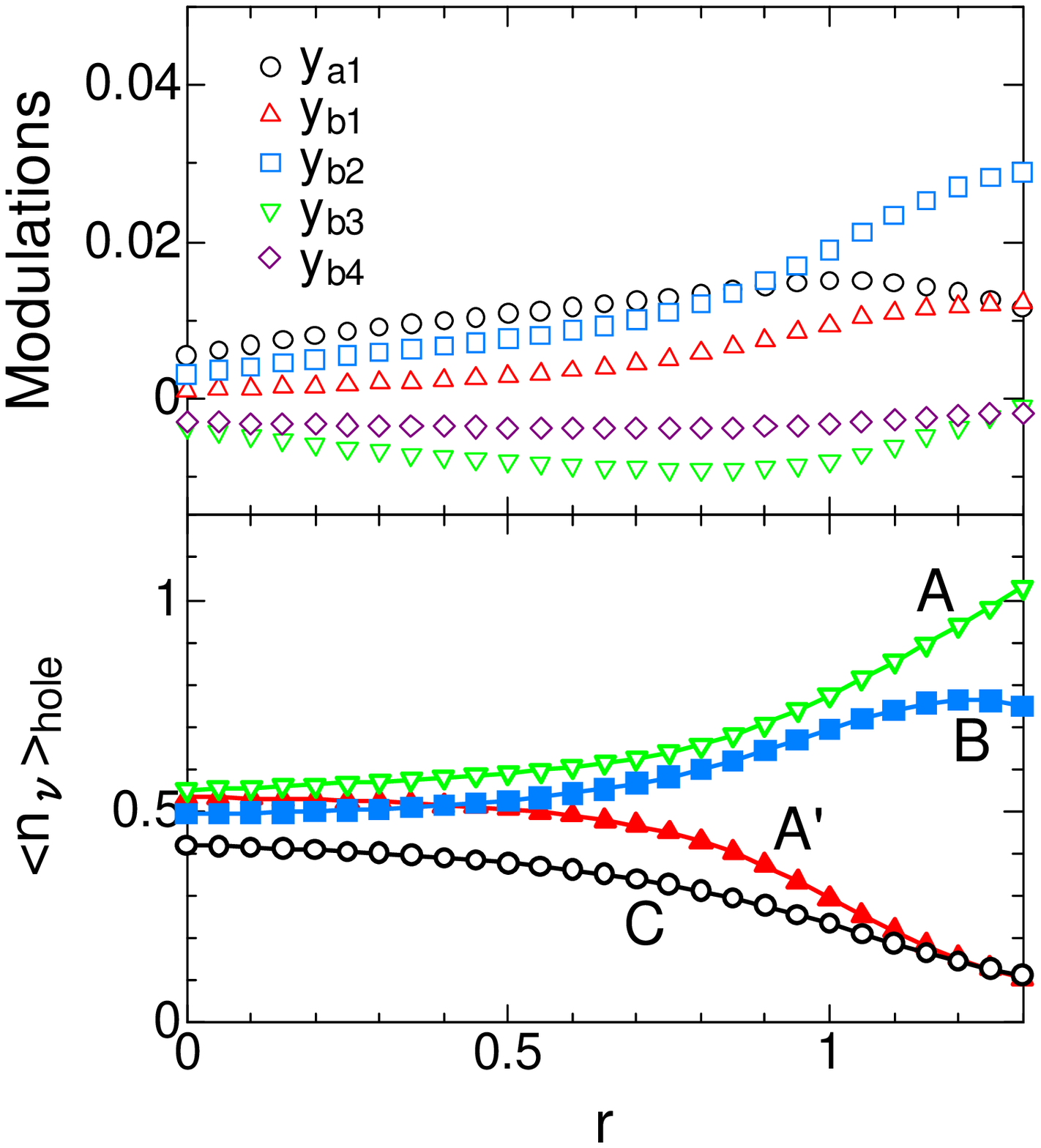}}
 \end{center}
\caption{(Color online) 
Modulations of transfer integrals (upper panel) and hole densities (lower panel) as a function of $r$ for $s_{a1}=s_{b1}=s_{b2}=s_{b3}=s_{b4}=0.05$ with $U$=0.7, $V_c$=0.35$r$, and $V_p$=0.30$r$. 
}
\label{f5}
\end{figure}
Finally, we study the effect of modulated intersite Coulomb interactions 
on the lattice distortion. 
By considering a situation with expansion or compression induced by a change in temperature or by external pressure, we show the modulations of the transfer integrals and the hole-density distribution with varying intersite Coulomb interactions in Fig. \ref{f5}. 
The HCO-$t_{b2'}$-$t_{b3}$ state remains the most stable over a wide region up to at least $r=1.3$. 
The effect of $s_{b2}$ is largest in stabilizing the HCO-$t_{b2'}$-$t_{b3}$ state for $\alpha$-I$_3$ because $y_{b2}$ is largest near $r=1$, which is a realistic value for $\alpha$-I$_3$. 
In the upper panel of Fig. $\ref{f5}$, $y_{b3}$ and $y_{b4}$ are shown to be negative. 
This inconsistency with the experimental results implies that the present model with transfer-modulating phonons is insufficient to reproduce the precise structural deformation.
%
\section{Strong-Coupling Analysis}
\label{strong}
To clarify the reason why the HCO state is stabilized by the present electron-phonon couplings, we consider the strong-coupling regime. 
We assume that the electron-electron interactions are much stronger 
than the transfer energies and that the on-site Coulomb interaction is the strongest among them ($U{\geq}2V_c>2V_p \gg t_{\mu}$). 

In refs. \citen{Hotta1} and \citen{Hotta2}, Hotta and coworkers presented another strong-coupling analysis. 
They focused on the charge degrees of freedom and adopted spinless fermions on an anisotropic triangular lattice. 
It should be noted that the spinless fermions correspond to the fully polarized spin state and the effects of intersite Coulomb interactions on the charge order can be considerably different from those in the singlet case, as shown in ref. \citen{Clay2}. 
In the following, we take both the spin and charge degrees of freedom into account. 

In the undistorted state of $\alpha$-I$_3$, charge disproportionation coexists with the inversion center because of the lower structural symmetry than that in the $\theta$-type salts.\cite{Kino1,Kobayashi1,Kobayashi2} 
In this state, the hole densities at sites $A$ and $A'$ remain equal, but that at site $B$ is rich, while that at site $C$ is poor [Fig. \ref{f1}(a)]. 
In the presence of lattice distortions, the inversion center disappears, and the hole densities at sites $A$ and $A'$ become different. 
We assume here that the holes are localized along the HCO-$t_{b2'}$-$t_{b3}$ chain at the strong-coupling limit, where the second-order correction to the energy and the elastic energy per unit cell is expected in terms of $hole$ transfer energies, $t_{\mu}^h$, 
\begin{eqnarray}
E_{\alpha}^{b2} &=& 2V_p 
        - \frac{ {t^h_{a1}}^2+{t^h_{a1'}}^2+{t^h_{a2}}^2+{t^h_{a3}}^2 }
               { V_c }
        - \frac{ {t^h_{b1'}}^2+{t^h_{b2}}^2+{t^h_{b3'}}^2+{t^h_{b4}}^2 }
               { 2V_c-V_p }
\nonumber \\
    &&  
        + \frac{ 4{t^h_{b2'}}^2 }{ U-V_p }
          \left\{
           \langle {\bf S}^{b2}_A \cdot {\bf S}^{b2}_B \rangle 
           - \frac{1}{4}
          \right\} 
        + \frac{ 4{t^h_{b3}}^2 }{ U-V_p }
          \left\{
           \langle {\bf S}^{b2}_B \cdot {\bf S}^{b2}_A \rangle 
           - \frac{1}{4}
          \right\} 
\nonumber \\
    &&  
        + \sum_{\mu} \frac{N_{\mu}}{ 2s_{\mu} } {y_{\mu}}^2
\;, 
\label{alpha-b2}
\end{eqnarray}
where $t^h_{\mu}=-t_{\mu}$ according to the electron-hole transformation and ${\bf S}^{b2}_{\nu}$ is the $S=1/2$ spin operator at site $\nu$ along the HCO-$t_{b2'}$-$t_{b3}$ chain. 
$N_{\mu}=2$ is the number of modulated pairs on the $\mu$ bonds in the unit cell. 
The second and third terms in eq. (\ref{alpha-b2}) are due to ``charge fluctuations'' between hole-rich and hole-poor sites. 
On the other hand, the fourth and fifth terms correspond to ``spin fluctuations'' on bonds between hole-rich sites. 

We expand eq. (\ref{alpha-b2}) using the relations 
shown in Table \ref{tbl1}, 
\begin{eqnarray}
E_{\alpha}^{b2} - 2V_p &=&
        E_{\alpha}^{\rm HT} 
        - \frac{2}{2V_c-V_p} 
           ( t_{b1}^{\rm HT} y_{b1} - t_{b2}^{\rm HT} y_{b2}
            -t_{b3}^{\rm HT} y_{b3} - t_{b4}^{\rm HT} y_{b4})
\nonumber \\
    &&
        - \frac{2}{V_c} {y_{a1}}^2
        - \frac{1}{2V_c-V_p} 
           ({y_{b1}}^2 + {y_{b2}}^2 + {y_{b3}}^2 + {y_{b4}}^2)
\nonumber \\
    &&
        + \frac{4}{U-V_p}
          \left\{
           \langle {\bf S}^{b2}_{A} \cdot {\bf S}^{b2}_{B} \rangle 
           - \frac{1}{4}
          \right\} 
          (2t_{b2}^{\rm HT}y_{b2} + {y_{b2}}^2)
\nonumber \\
    &&
        + \frac{4}{U-V_p}
          \left\{
           \langle {\bf S}^{b2}_{B} \cdot {\bf S}^{b2}_{A} \rangle 
           - \frac{1}{4}
          \right\} 
          (2t_{b3}^{\rm HT}y_{b3} + {y_{b3}}^2)
\nonumber \\
    &&  
        + \sum_{\mu} \frac{N_{\mu}}{ 2s_{\mu} } {y_{\mu}}^2 
\;, 
\label{alpha-b2-2}
\end{eqnarray}
where
\begin{eqnarray}
E_{\alpha}^{\rm HT} &=& 
        - \frac{1}{V_c} 
           ( 2{t_{a1}^{\rm HT}}^2 
            + {t_{a2}^{\rm HT}}^2 + {t_{a3}^{\rm HT}}^2)
        - \frac{1}{2V_c-V_p} 
           ( {t_{b1}^{\rm HT}}^2 + {t_{b2}^{\rm HT}}^2 
            +{t_{b3}^{\rm HT}}^2 + {t_{b4}^{\rm HT}}^2)
\nonumber \\
    &&
        + \frac{4 {t_{b2}^{\rm HT}}^2}{U-V_p}
          \left\{
           \langle {\bf S}^{b2}_{A} \cdot {\bf S}^{b2}_{B} \rangle 
           - \frac{1}{4}
          \right\}
        + \frac{4 {t_{b3}^{\rm HT}}^2}{U-V_p}
          \left\{
           \langle {\bf S}^{b2}_{B} \cdot {\bf S}^{b2}_{A} \rangle 
           - \frac{1}{4}
          \right\}
\end{eqnarray}
is the second-order correction to the energy per unit cell in the undistorted state. 
We define the energy difference as a function of $y_{b2}$ as, 
\begin{eqnarray}
D_{\alpha}^{b2} &=& E_{\alpha}^{b2}(y_{b2})-E_{\alpha}^{b2}(0)
\nonumber \\
&=& 
  \frac{1}{2V_c-V_p} \left(
                      2 t_{b2}^{\rm HT} y_{b2} - {y_{b2}}^2
                     \right)
+ \frac{4}{U-V_p} \left\{ 
                  \langle {\bf S}^{b2}_A \cdot {\bf S}^{b2}_B \rangle 
                         - \frac{1}{4}
                  \right\}
                  \left(
                   2 t_{b2}^{\rm HT} y_{b2} + {y_{b2}}^2
                  \right)
\nonumber \\
&&
     + \frac{N_{b2}}{2 s_{b2}} {y_{b2}}^2
\;. 
\label{d-alpha-b2}
\end{eqnarray}
Because $t_{b2}^{\rm HT}>0$, if $t_{b2}<t_{b2'}$ results from lattice distortions, namely, $y_{b2}>0$, the first and second terms of eq. (\ref{d-alpha-b2}) represent the energy loss from charge fluctuations and the energy gain from spin fluctuations, respectively. 
On the other hand, if $y_{b2}<0$, the total energy is lowered by charge fluctuations and raised by spin fluctuations. 
In the distorted state of $\alpha$-I$_3$, the energy gain from spin fluctuations dominates, so that the former situation ($y_{b2}>0$) is realized. 
In other words, if spin or charge fluctuations occur alone, they always lower the energy. 
When they compete (spin fluctuations favor $y_{b2}>0$, while charge fluctuations favor $y_{b2}<0$), cancellation occurs as in the present case. 

From eq. (\ref{d-alpha-b2}), we can easily derive the following relation: 
\begin{eqnarray}
y_{b2} &=& \frac{(\chi_s^{\alpha}+\chi_c^{\alpha}) t_{b2}^{\rm HT}}
                {N_{b2} - (\chi_s^{\alpha}-\chi_c^{\alpha}) s_{b2}}
                s_{b2}
\nonumber\\
       &\rightarrow&
           \frac{(\chi_s^{\alpha}+\chi_c^{\alpha}) t_{b2}^{\rm HT}}
                {N_{b2}}
                s_{b2}
           \ \ \ \ \ \ 
           ({\rm as}\; s_{b2} \rightarrow 0)
\;, 
\label{yb2-sb2}
\end{eqnarray}
where 
\begin{eqnarray}
\begin{array}{lr}
\displaystyle
\chi_c^{\alpha} = -\frac{2}{2V_c-V_p}<0, &
\displaystyle
\chi_s^{\alpha} = -\frac{8}{U-V_p}
                   \left(
                    \langle
                     {\bf S}^{b2}_A \cdot {\bf S}^{b2}_B
                    \rangle 
                    - \frac{1}{4}
                   \right)>0
\;. 
\end{array}
\label{alpha-X}
\end{eqnarray}
As explained above, the linear coefficient in eq. (\ref{yb2-sb2}) depends on both spin fluctuations ($\chi_s^{\alpha}$) and charge fluctuations ($\chi_c^{\alpha}$), which partially cancel each other, resulting in the smallness of lattice distortions. 
Because of this, the lattice effect on $\alpha$-I$_3$ is much smaller than that on $\theta$-RbZn shown in the Appendix. 
To compare the strong-coupling analysis with the numerical results, the linear increase of the modulations are shown in the upper panels of Figs. \ref{f2} and \ref{f3}. 
For the values of the nearest-neighbor spin correlations, we use $\langle{\bf S}^{b2}_{A}\cdot{\bf S}^{b2}_{B}\rangle{\simeq}-0.74$ and $\langle{\bf S}^{b2}_{B}\cdot{\bf S}^{b2}_{A}\rangle{\simeq}-0.04$ on the bonds with the larger and smaller exchange couplings, $J=4{t^{\rm HT}_{b2}}^2/(U-V_p)$ and $J'=4{t^{\rm HT}_{b3}}^2/(U-V_p)$, respectively. 
These are obtained from the exact-diagonalization calculation of the 6-site $S=1/2$ bond-alternating spin chain with couplings $J$ and $J'$, respectively, which are included in the 12-site cluster used in Figs. \ref{f2} and \ref{f3}. 
Accordingly, $|\chi_s^{\alpha}|=|-8{\times}(-0.74-0.25)/0.35| \sim 22.6$ is larger than $|\chi_c^{\alpha}|=|-2/0.35| \sim 5.7$. 
These analytic results are qualitatively consistent with the numerical results. Of course, at the limit of strong interactions ($U$, $V_c$ and $V_p$ $\rightarrow$ $\infty$), the numerical results coincide with the analytic results (not shown).

\section{Summary}
Using the 2D 3/4-filled extended Hubbard model for an anisotropic triangular lattice, we have investigated the effects of spin and charge fluctuations and electron-phonon couplings on the charge order in $\alpha$-I$_3$. 
The hole densities and the modulations of transfer integrals are 
obtained by the exact-diagonalization method. 

Owing to the low-symmetry structure, charge disproportionation is known to exist even without lattice distortion, as observed in the high-temperature phase. 
The horizontal-stripe charge-ordered phase is induced by electron-phonon couplings in the present small clusters. 
Thus, the molecular displacements assist the formation of the horizontal-stripe CO, as experimentally observed.\cite{Kakiuchi} 
The effect of modulations is largest on the $b2$ bonds and is significant in stabilizing the HCO-$t_{b2'}$-$t_{b3}$ state since nearly isotropic long-range Coulomb interactions do not favor any particular CO pattern in the present small-cluster calculation. 

Strong-coupling analysis is used to elucidate the mechanism for stabilizing the HCO state, and the $y_{b2}$ modulations increase linearly with respect to the electron-phonon coupling $s_{b2}$. 
It is clarified that the effect of spin fluctuations is partially canceled by charge fluctuations. 
All the present results are consistent with the mean-field results for the same model,\cite{Tanaka1,Tanaka2} although they show the horizontal-stripe CO even 
without electron-phonon couplings at the thermodynamic limit. 
See also ref. \citen{Tanaka2} for the lattice effects on the threefold state, which we do not consider here because it is not the lowest-energy state.

\section*{Acknowledgments}
The authors are grateful to S. Iwai, T. Kakiuchi, and H. Sawa for showing their data prior to publication, and they would like to thank Y. Tanaka and Y. Yamashita for fruitful discussions. 
This work was supported by the Next-Generation Supercomputer Project (Integrated Nanoscience) and Grants-in-Aid from the Ministry of Education, Culture, Sports, Science and Technology, Japan. 

\appendix
\section{The Case of $\theta$-(BEDT-TTF)$_2$RbZn(SCN)$_4$}
We compare the effects of spin and charge fluctuations in $\theta$-RbZn with 
those in $\alpha$-I$_3$. 
It was clarified in our previous paper\cite{Miyashita} that structural deformations, in particular, molecular rotations assist the formation of the horizontal-stripe CO in $\theta$-RbZn, which has been experimentally observed.\cite{Watanabe} 
The mechanism for stabilizing the HCO state and the linearly increasing modulation are explained by strong-coupling perturbation theory.\cite{Miyashita} 
Here, we expand the transfer energies to clarify these points. 

First, for the $\theta$-RbZn salt, we define the transfer integrals for electrons as 
\begin{eqnarray}
\begin{array}{ll}
t_{c1}=t_{c}^{\rm HT}-y_{c}, &
t_{c2}=t_{c}^{\rm HT}+y_{c}, \\
t_{p1}=t_{p1}^{\rm HT}+y_{p1}, &
t_{p3}=t_{p1}^{\rm HT}-y_{p1}, \\
t_{p2}=t_{p4}^{\rm HT}+y_{p4}, &
t_{p4}=t_{p4}^{\rm HT}-y_{p4},
\end{array}
\label{theta-transfer}
\end{eqnarray}
where $0<t_{c}^{\rm HT}<t_{p1}^{\rm HT}=-t_{p4}^{\rm HT}$,\cite{Miyashita,Tanaka1,Tanaka2} and the modulations $y_{c(a)}$ and $y_{p4}$ originate from the molecular translations along the $c(a)$-direction and the molecular rotations, respectively. 

When the holes are localized on the $p4$ bonds and the HCO-$t_{p4}$ is realized (see Fig. 1 of ref. \citen{Miyashita}), the second-order correction to the energy and the elastic energy per unit cell is given for the holes by 
\begin{eqnarray}
E_{\theta}^{p4} &=& 2V_p 
        - \frac{ 2({t^h_{c1}}^2+{t^h_{c2}}^2) }{ V_c }
        - \frac{ 2({t^h_{p1}}^2+{t^h_{p3}}^2) }{ 2V_c-V_p }
\nonumber \\
    &&  
        + \frac{ 8{t^h_{p4}}^2 }{ U-V_p }
          \left\{
           \langle {\bf S}^{p4}_1 \cdot {\bf S}^{p4}_4 \rangle - \frac{1}{4}
          \right\} 
        + \sum_{\mu} \frac{N_{\mu}}{ 2s_{\mu} } {y_{\mu}}^2 
\;, 
\label{theta-p4}
\end{eqnarray}
where $t^h_{\mu}=-t_{\mu}$ due to the electron-hole transformation and ${\bf S}^{p4}_{1(4)}$ denotes the $S=1/2$ spin operator at site 1(4) on the $p4$ bond. 
$N_{\mu}=4$ is the number of modulated pairs on the $\mu$ bonds in the unit cell. 
At the strong-coupling limit, a 1D half-filled chain is formed on the $p4$ bonds, and $\langle {\bf S}^{p4}_1 \cdot {\bf S}^{p4}_4 \rangle$=
$-\ln 2 + 1/4 \simeq -0.443$ is the exact ground-state energy of the isotropic $S=1/2$ Heisenberg chain. 

Similarly to the case of $\alpha$-I$_3$, eq. (\ref{theta-p4}) is expanded using the relations in eq. (\ref{theta-transfer}),
\begin{eqnarray}
E_{\theta}^{p4} - 2V_p&=& E_{\theta}^{\rm HT} 
        - \frac{4}{V_c}
           {y_c}^2
        - \frac{4}{2V_c-V_p}
           {y_{p1}}^2
\nonumber \\
    &&  
        + \frac{ 8(-2t_{p4}^{\rm HT}y_{p4}+{y_{p4}}^2) }
               { U-V_p }
          \left\{
           \langle {\bf S}^{p4}_1 \cdot {\bf S}^{p4}_4 \rangle - \frac{1}{4}
          \right\} 
        + \sum_{\mu} \frac{N_{\mu}}{ 2s_{\mu} } {y_{\mu}}^2 
\;, 
\label{theta-p4-2}
\end{eqnarray}
where $E_{\theta}^{\rm HT}$ is the second-order correction for the undistorted state given by
\begin{eqnarray}
E_{\theta}^{\rm HT} = 
        - \frac{4}{V_c}
           {t_c^{\rm HT}}^2
        - \frac{4}{2V_c-V_p}
           {t_{p1}^{\rm HT}}^2
        + \frac{ 8{t_{p4}^{\rm HT}}^2}{ U-V_p }
          \left\{
           \langle {\bf S}^{p4}_1 \cdot {\bf S}^{p4}_4 \rangle - \frac{1}{4}
          \right\} 
\;. 
\label{theta-HT}
\end{eqnarray}
The HCO-$t_{p4}$ state is realized in the low-temperature phase of $\theta$-RbZn because the total energy gain originating from spin fluctuations is large. 
Such uniform enhancement of the $p4$ bonds is allowed by the molecular rotation $y_{p4}$,\cite{Miyashita,Tanaka1} 
which would be impossible if only molecular translations were present. 

From eq. (\ref{theta-p4-2}), the relation between $y_{p4}$ and $s_{p4}$ is given by 
\begin{eqnarray}
y_{p4} &=& -\frac{2\chi_s^{\theta} t_{p4}^{\rm HT}}
                 {N_{p4 - }2\chi_s^{\theta} s_{p4}}
                s_{p4}
\nonumber\\
       &\rightarrow&
           -\frac{2\chi_s^{\theta} t_{p4}^{\rm HT}}
                 {N_{p4}}
                s_{p4}
           \ \ \ \ \ \ 
           ({\rm as}\; s_{p4} \rightarrow 0)
\;, 
\label{yp4-sp4}
\end{eqnarray}
where 
\begin{eqnarray}
\chi_s^{\theta} = -\frac{8}{U-V_p}
                   \left\{
                    \langle
                     {\bf S}^{p4}_1 \cdot {\bf S}^{p4}_4
                    \rangle 
                    - \frac{1}{4}
                   \right\}>0
\;. 
\label{theta-X}
\end{eqnarray}
Equation (\ref{theta-X}) shows that once the HCO-$t_{p4}$ is stabilized, the total energy is lowered by the positive $y_{p4}$. 
This is also achieved in the realistic case of $U$=0.7, $V_c$=0.35, and $V_p$=0.30. 
It should be noted that the linear coefficient in eq. (\ref{theta-X}) depends $only$ on spin fluctuations ($\chi_s^{\theta}$), in contrast to the case of $\alpha$-I$_3$. 
In $\theta$-RbZn, spin fluctuations induce large lattice distortions in the form of molecular rotation, and the total energy is markedly lowered. 
It should be noted that the distortion $y_c$ is explained by third-order perturbation theory.\cite{Miyashita}

\end{document}